\documentclass[aps,preprint,showpacs]{revtex4}
\usepackage{amsfonts}
\usepackage{amsmath}
\usepackage{amssymb}
\usepackage{graphicx}

\begin{document}
\preprint{ }
\title{Field induced resistivity anisotropy in SrRuO$_{3}$ films}
\author{Yishai Shperber$^{1}$}
\author{Isaschar Genish$^{1}$}
\author{James W. Reiner$^{2}$}
\author{Lior Klein$^{1}$}
\affiliation{$^{1}$Department of Physics, Nano-magnetism Research
Center, Institute of Nanotechnology and Advanced Materials,
Bar-Ilan
University, Ramat-Gan 52900, Israel}
\affiliation{$^{2}$Department of Applied Physics,
Yale University, New Haven, Connecticut 06520-8284}

\begin{abstract}
SrRuO$_{3}$ is an itinerant ferromagnet with orthorhombic structure and
uniaxial magnetocrystalline anisotropy - features expected to
yield resistivity anisotropy. Here we explore changes in the resistivity anisotropy of epitaxial SrRuO$_{3}$ films
due to induced magnetization in the paramagnetic state by using the planar Hall effect. We find that the effect of the induced magnetization on the in-plane anisotropy is strongly angular dependent and we provide a full description of this behavior at 160 K for induced magnetization in the (001) plane.
\end{abstract}






\maketitle

The itinerant ferromagnet SrRuO$_{3}$ has attracted considerable
experimental and theoretical effort for its intriguing properties
\cite{trans, nonfermidisorder, optical, Matthiessen's}.
Being almost cubic, thin films of SrRuO$_{3}$
exhibit rather small anisotropy which is difficult to characterize
by comparing resistivity measurements taken on patterns with
current flowing along different directions relative the
crystallographic axes. Consequently, we have determined the zero field
anisotropy of epitaxial films of this compound, by measuring the planar Hall effect (PHE)
\cite{Genish1}, a method which provides a good local measurement of the anisotropy with high accuracy.
To explore the contribution of the magnetization $\textbf{M}$ to the anisotropy we examine
how changing the magnitude and orientation of the magnetization
affects the resistivity anisotropy.
We find that the anisotropy is proportional to M$^{2}$ for any angle in the (001) plane; however, the proportionality coefficient is strongly angular dependent.

Our samples are epitaxial films of SrRuO$_{3}$ grown on slightly
miscut (2$^{\circ}$) SrTiO$_{3}$. The films are orthorhombic
($a=5.53$ \AA , $b=5.57$ \AA , $c=7.85$ \AA) \cite{Marshall} and
their Curie temperature is $\sim150$ K. The films grow with the
in-plane $c$ axis perpendicular to the miscut, and $a$ and $b$
axes at 45$^{\circ}$ relative to the film plane. The films exhibit
uniaxial magnetocrystalline anisotropy with the easy axis along
the $b$ axis at T$\geq$$T_{c}$ \cite{Marshall}.

The data presented here are from a 27 nm thick film with
resistivity ratio of $\sim12$. The measurements were done at 160 K on a pattern with current direction along [1$\bar{1}$1] for which the PHE attains its maximal value \cite{Genish1}.

\begin{figure}[ptb]
\includegraphics[scale=0.45, trim=50 150 0 50]{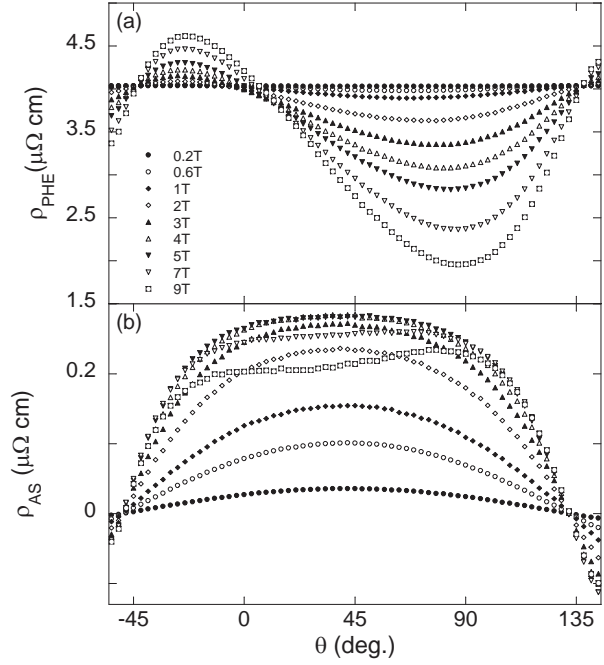}
\caption{$\rho_{PHE}$ (a) and $\rho_{AS}$ (b) vs $\theta$, the angle between the applied field and the film normal.}
 \label{PHEandASimvsTetha}
\end{figure}

Figure~$\ref{PHEandASimvsTetha}$(a) and Figure~$\ref{PHEandASimvsTetha}$(b) show the transverse signal (symmetric (a) and antisymmetric (b)) vs $\theta$, the angle between the applied magnetic field and the normal to the film. Positive $\theta$ corresponds to an anticlockwise rotation. The PHE is the symmetric  part of the transverse signal \cite{PHE1, PHE2, amr} and in our configuration  $\rho_{PHE}=\rho_{[1\bar{1}0]}-\rho_{[001]}$ where $\rho_{[1\bar{1}0]}$ is the longitudinal resistivity along the $[1\bar{1}0]$ direction and
$\rho_{[001]}$ is the longitudinal resistivity along $[001]$.
The antisymmetric part ($\rho_{AS}$) consists of the ordinary Hall effect (OHE) and the extraordinary Hall effect (EHE) \cite{EHE}.

\begin{figure}[ptb]
\includegraphics[scale=0.50, trim=50 -50 0 0]{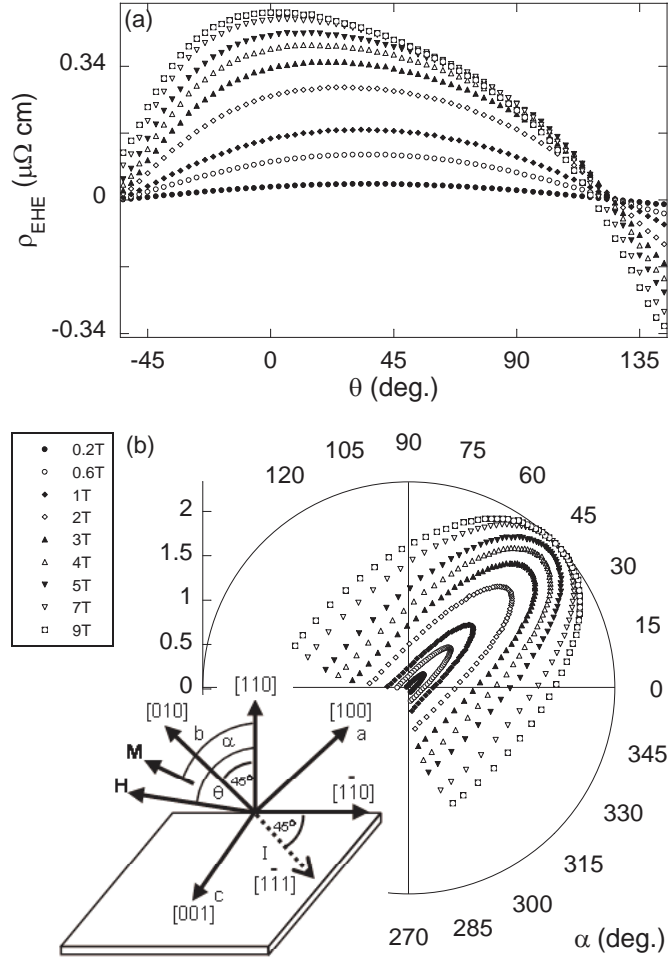}
\caption{a) The EHE part of the antisymmetric signal vs $\theta$ the angle between the applied field and the film normal. b) The magnitude (arbitrary units) and direction of the magnetization obtained for rotating different fields in the (001) plane. Inset: Crystallographic directions of the film and definitions of the angles $\theta$ and $\alpha$.}
 \label{polar4}
\end{figure}

Subtracting the OHE contribution from $\rho_{AS}$ yields the EHE which we present in Figure~$\ref{polar4}$(a). As is well known, the EHE is proportional to $M_{\perp}$, the component of $\mathbf{M}$ which is perpendicular to the plane of the film. However, the magnetocrystalline anisotropy of SrRuO$_{3}$ allows us to determine the full vector $\mathbf{M}$. As the easy
axis is at 45$^{\circ}$ to the film normal, a field $\textbf{H}$, applied
in the (001) plane at angle $\theta$ relative to the normal of the sample's plane,
creates magnetization $\textbf{M}$ at angle $\alpha$ (Figure~$\ref{polar4}$(b)-inset), where
the same field at angle $90-\theta$ creates $\textbf{M}$ with the same magnitude but at angle $90-\alpha$.
Consequently, by measuring the $\rho_{EHE}$ at angle $\theta$ $(\rho_{EHE}(\theta))$ and 90-$\theta$ $(\rho_{EHE}(90-\theta))$ we obtain
 $\rho_{EHE}(\theta)\propto$$M_{\perp}(\theta)$
and $\rho_{EHE}(90-\theta)\propto$$M_{\parallel}(\theta)$, where $M_{\parallel}$ and $M_{\perp}$ are the in-plane and
perpendicular components of $\textbf{M}$, respectively.
Thus we obtain a full description of the magnetization vector \cite{Genish2}.

Figure~$\ref{polar4}$(b) shows the vector of magnetization in arbitrary units obtained for some of the fields we used in our measurements. In this polar graph we see that the magnetization has its maximal value when it is aligned along the easy axis and its minimal value when it lies along the hard axis. The data indicate that with $\mathbf{H}=1$ T the induced magnetization along the easy axis is larger than the induced magnetization along the hard axis by a factor of $\sim5.5$, in good agreement with a previous report \cite {Yevgeny}.

\begin{figure}[ptb]
\includegraphics[scale=0.50, trim=50 200 0 230]{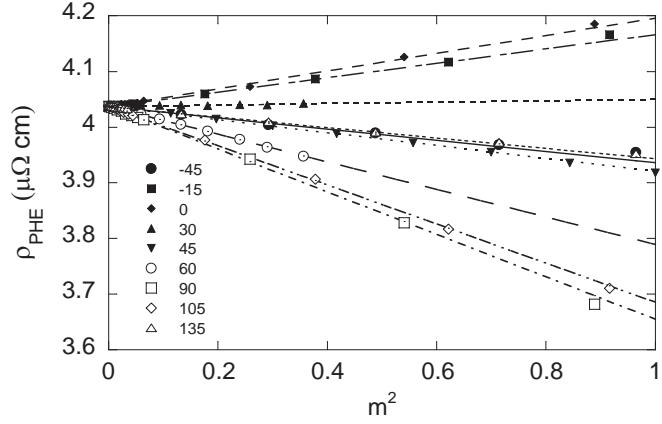}
\caption{$\rho_{PHE}$ vs $m^{2}$. The lines are linear fits .}
 \label{PHEvsM2}
\end{figure}

\begin{figure}[ptb]
\includegraphics[scale=0.50, trim=50 200 0 250]{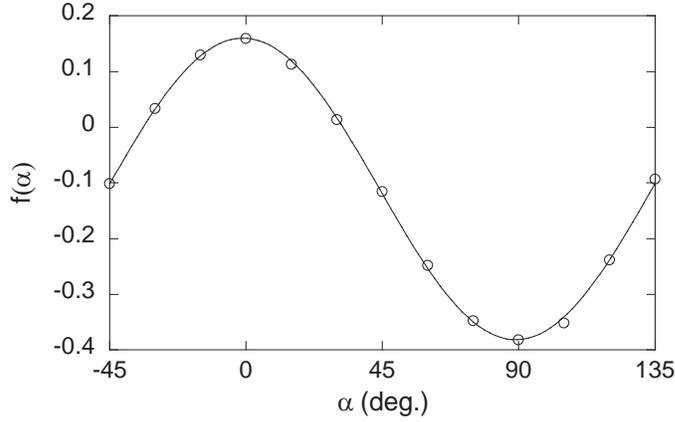}
\caption{$f(\alpha)$ where $\alpha$ is the angle between $\mathbf{m}$ and the film normal. The solid line is a fit to Eq.~\ref{2}.}
 \label{AvsAlpha}
\end{figure}

Figure~$\ref{PHEvsM2}$ shows the dependence of $\rho_{PHE}$ on m$^{2}$ where $\textbf{m}$ is defined as $\mathbf{M}/|\widetilde{\mathbf{M}}|$ and $\widetilde{\mathbf{M}}$ is the magnetization obtained with a field of 1 T applied along the easy axis. Based on the construction of the vector of magnetization presented in Figure~$\ref{polar4}$, we can extract the change in $\rho_{PHE}$ as a function of $\mathbf{m}$ for a given orientation. We clearly see that:

\begin{equation}
\\\rho_{PHE}=\rho^{o}_{PHE}+f(\alpha)m^{2} \label{1}
\end{equation}

where $\rho^{0}_{PHE}$ is the PHE measured at zero field and $f(\alpha)$ is the coefficient of m$^{2}$ when $\mathbf{m}$ is along $\alpha$. Figure~$\ref{AvsAlpha}$ shows f($\alpha$) for some of the $\alpha$'s. We see that there is an excellent fit of f($\alpha$) with:

\begin{equation}
\\f(\alpha)=a+bsin^{2}\alpha \label{2}
\end{equation}

This means that one can write $\rho_{PHE}$ as:

\begin{equation}
\\\rho_{PHE}=\rho^{0}_{PHE}+am^{2}+bm_{\parallel}^{2} \label{3}
\end{equation}

where a=0.16 and b=-0.54.

The dependence we have found indicate that there are two effects of the induced magnetization on the resistivity anisotropy. First, there is a contribution independent of the direction of $\mathbf{M}$ with a positive coefficient $\textit{a}$ which means that the anisotropy increases with increasing magnetization. Second, there is another contribution sensitive to the in-plane projection of the magnetization with a negative coefficient $\textit{b}$ which means that the anisotropy decreases with increasing magnetization. The extremal points of $f(\alpha)$ are at $\alpha=0^{\circ}$, $90^{\circ}$. When $\alpha=0^{\circ}$ the induced magnetization is normal to J along both $[1\bar{1}0]$ and [001]. At this angle the anisotropy contribution is minimal and $f(\alpha)$ attains its maximal positive value. When $\alpha=90^{\circ}$, the magnetization is parallel to $[1\bar{1}0]$ and perpendicular to [001] and therefore its anisotropic contribution is maximal and $f(\alpha)$ attains its maximum negative value.

L. K. acknowledges support by the Israel Science Foundation
founded by the Israel Academy of Sciences and Humanities. J. W. R. grew the samples at Stanford University in the laboratory of M. R. Beasley.

\end{document}